\documentclass[preprint,10pt]{elsarticle}
\usepackage{graphicx}
\usepackage[version=3]{mhchem}
\usepackage{float}
\usepackage{siunitx}
\usepackage[hidelinks]{hyperref}
\usepackage{amsmath,amsthm,amssymb}
\usepackage{natbib}
\usepackage{fleqn,txfonts}
\usepackage{xcolor}

\usepackage{xspace}

\newcommand{\ket} [1] {\ensuremath{{|} {#1} \rangle}}

\newcommand{\bra} [1] {\ensuremath{\langle {{#1}} |}}

\newcommand{\beq} {\begin{equation}}
\newcommand{\eeq} {\end{equation}}
\newcommand{\cpara} {\ensuremath{{c\parallel B_0}}\xspace}
\newcommand{\cperp} {\ensuremath{{c\bot B_0}}\xspace}
\journal{Journal of Magnetic Resonance}

\begin{document}

\begin{frontmatter}

\title{Unusual $^{209}$Bi NMR quadrupole effects in topological insulator \ce{Bi2Se3}}


\author[1,2,3]{Robin Guehne}
\author[4]{Vojt\v{e}ch Chlan}
\author[2]{Grant V. M. Williams}
\author[2,3]{Shen V. Chong}
\author[5]{Kazuo Kadowaki}
\author[1]{Andreas P\"oppl}
\author[1]{J\"{u}rgen Haase\corref{cor1}}

\cortext[cor1]{Corresponding author}

\address[1]{Felix Bloch Institute for Solid State Physics, University of Leipzig, Linnéstrasse 5, 04103 Leipzig, Germany}
\address[2]{The MacDiarmid Institute for Advanced Materials and Nanotechnology, SCPS, Victoria University of Wellington, PO Box 600, Wellington 6140, New Zealand}
\address[3]{Robinson Research Institute, Victoria University of Wellington, PO Box 33436, Lower Hutt 5046, New Zealand}
\address[4]{Charles University in Prague, Faculty of Mathematics and Physics, V Holešovičkách 2, 180 00, Prague 8, Czech Republic}
\address[5]{Division of Materials Science, Faculty of Pure and Applied Sciences, University of Tsukuba, 1-1-1, Tennodai, Tsukuba, Ibaraki 305-8573, Japan}

\begin{abstract}
Three-dimensional topological insulators are an important class of modern materials, and a strong spin-orbit coupling is involved in making the bulk electronic states very different from those near the surface. Bi$_2$Se$_3$ is a model compound, and $^{209}$Bi NMR is employed here to investigate the bulk properties of the material with focus on the quadrupole splitting. It will be shown that this splitting measures the energy band inversion induced by spin-orbit coupling in quantitative agreement with first-principle calculations. Furthermore, this quadrupole interaction is very unusual as it can show essentially no angular dependence, e.g., even at the magic angle the first-order splitting remains. Therefore, it is proposed that the magnetic field direction is involved in setting the quantization axis for the electrons, and that their life time leads to a new electronically driven relaxation mechanism, in particular for quadrupolar nuclei like $^{209}$Bi. While a quantitative understanding of these effects cannot be given, the results implicate that NMR can become a powerful tool for the investigation of such systems.
\end{abstract}

\begin{keyword}
Energy band inversion \sep Spin-orbit coupling \sep Quadrupole interaction \sep Topological insulator



\end{keyword}

\end{frontmatter}

\section{Introduction}

Ten years ago, it was predicted that \ce{Bi2Se3} is a topological insulator \cite{Zhang2009}, i.e., in addition to the conventional insulating bulk it has a conducting surface. The corresponding metallic surface states are a consequence of strong spin-orbit coupling that leads to an inversion of bands (and atomic states) with different symmetries \cite{Zhang2009}. There is nothing new in terms of chemistry about \ce{Bi2Se3}, but these new electronic states are special since they are protected by topology. This may not only be of interest in physics, but also chemistry, e.g., if one thinks of heterogeneous catalysis.

For those familiar with reciprocal space and general symmetry arguments, the understanding of the concept of topological insulators is perhaps simple. However, NMR practitioners or those who are used to local thinking may experience difficulties in imagining real space consequences for these materials. In fact, it is not evident whether a bulk, local probe like NMR has something important to say about topological insulators. The characterization of the few nuclei near the material's surface, which must be affected by the special surface electrons, is a different challenge. Nanometer-size grains with large surface to volume ratio, required for such investigations, may have different bulk properties, as well. Therefore, the bulk NMR properties of larger crystals should be understood first, an effort we pursue.

Before we address the very few NMR papers (with controversial observations), we would like to introduce a simple chemical example with consequences in topology, which we found helpful in thinking about such systems.

\begin{figure*}
\centering
\includegraphics[width=8cm]{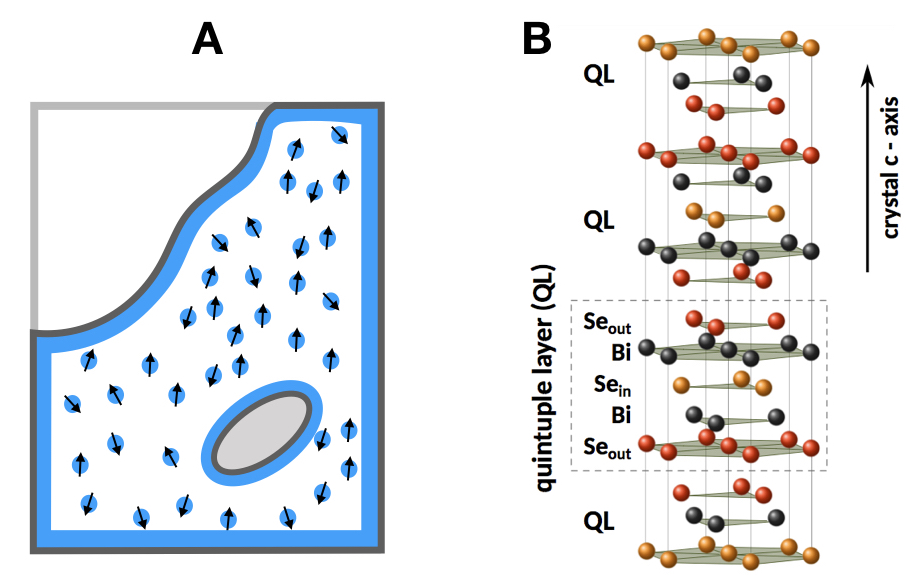}
\caption{\label{fig:water} A, water in a closed container at given temperature can hold two different types of molecules those in the liquid (blue full lines), or in the gas (blue circles with arrows) in particular total angular momentum ($J$) states. This scenario is robust and independent of the shape of inner or outer surfaces. B, schematic crystal structure of \ce{Bi2Se3}; parallel planes (with Bi or Se atoms) stacked in crystal $c$-direction are of fcc type so that three quintuple layers (Se$_{\rm out}$-Bi-Se$_{\rm in}$-Bi-Se$_{\rm out}$ planes) are needed before the stacking repeats. The unit cell consists of two chemically equivalent Bi, one Se$_{\rm in}$, and two Se$_{\rm out}$ atoms. Any Bi$_2$Se$_3$ surface (i.e., an interface to vacuum or another regular insulator) carries topologically protected metallic states. }
\end{figure*}
We believe that simple water molecules can offer some clues. Without any interference as a free gas, individual water molecules will be found in particular total angular momentum states ($J = 0, 1, 2, ...$,  with consequences for its nuclear spins according to the Pauli principle). In astrophysics, those differences, detected at great distances, prove facts about the universe \cite{Emprechtinger2012}. However, in condensed matter systems water molecules loose some of their free gas identity due to interactions. Nevertheless, they can still have different reactivities, as recently reported \cite{Kilaj2018}, so that there might be even more surprises in some systems. NMR knows about this \cite{Meier2018}, but related effects are better studied for other small molecules, e.g., para and ortho molecular hydrogen \cite{Lipsicas1961,Abragam1961}. This special quantum behavior of small molecules also affects nuclear relaxation, e.g., the famous spin-rotation interaction between nuclei and the molecular magnetic moment related to $J$ was well studied \cite{Oppenheim1961} (and that may have its analogue in topological systems, as we argue below).

If one thinks of ideal liquid water, it can only exist if enclosed in a container where it will be in equilibrium with its free gas molecules. This co-existence of liquid and gas can be looked upon in terms of topology in the sense that the water molecules are different objects if in liquid versus gas (with some appropriate time and length scales). Then, a closed container with walls that favor the liquid state has some resemblance with a topological insulator. For the latter, surface electrons have different topological properties compared to those in the bulk. Note that such scenarios are independent of the shape of the surface, cf.~Fig.~\ref{fig:water}. And if there is another - appropriate - surface within the container (e.g., aerosol particles in a water atmosphere) it could be covered by a liquid film, as well. This may help comprehend why modifications of outside or inside surfaces of topological insulators are so interesting, and why a local probe like NMR can be of great help in the study of those effects.

In \ce{Bi2Se3} it is the strong spin-orbit coupling that disappears on the surface and changes the topology. Of course, any chemical (and subsequent electronic) surface reconstruction can complicate this simple scenario. Nevertheless, it appears from experimental surface probes that the predicted states are indeed present in this material \cite{Xia2009}.

Most of the reports on NMR of topological insulators, i.e., \ce{Bi2Se3} and \ce{Bi2Te3}, focussed on the spin $1/2$ nuclei $^{77}$Se and $^{125}$Te. They concerned the investigation of single crystals \cite{Podorozhkin2015,Georgieva2016,Matano2016,Antonenko2017a,Antonenko2017b}, powders \cite{Taylor2012,Koumoulis2014,Levin2016}, or nano-powders \cite{Koumoulis2013,Choi2018,Choi2018b}, and addressed relaxation, shifts, and the influence of surface states, but also the presence of a strong indirect nuclear spin coupling that gives rise to extensive NMR line broadening in \ce{Bi2Se3} \cite{Georgieva2016}. The understanding of NMR in strongly spin-orbit coupled systems is not well developed. For example, as was shown very recently, Knight shift and orbital shift in spin-orbit coupled metals have rather different phenomenology compared to conventional systems \cite{Boutin2016}.

There have been only a few accounts of $^{209}$Bi NMR of \ce{Bi2Se3} \cite{Young2012,Nisson2013,Nisson2014,Mukhopadhyay2015}. As a spin $9/2$ system, $^{209}$Bi appears to be affected by a smaller than expected quadrupole interaction of $\sim$\SI{150}{kHz} \cite{Young2012}. The overall lineshape was found to be unusual, i.e., the intensity pattern of the individual transitions shows the lowest intensity for the central transition, and the intensity grows towards the outermost satellites. Young et al. \cite{Young2012} pointed to rapid transverse relaxation ($T_2$) to be the origin of this intensity anomaly. Nisson et al. \cite{Nisson2014} measured the angular dependent spectra and found a significant loss of signal intensity when rotating the single crystal away from \cpara (the magnetic field $B_0$ parallel to the crystal $c$-axis). They found discrepancies when comparing averaged orientation dependent single crystal spectra with those obtained from microscopic and nanoscopic powders. It was suggested \cite{Georgieva2016} that the indirect nuclear spin coupling in these systems  is behind the rapid $T_2$. 

Here we will address the $^{209}$Bi NMR quadrupole splitting in more detail, and we will see that it can be highly unusual in these materials. We will show that the small quadrupole interaction is a consequence of the inverted band structure that affects the relative occupation of Se and Bi atomic levels. Thus, the electronic properties bear direct consequences for NMR, and there is even quantitative agreement with first-principle calculations. Even more stunning is the angular dependence of the Bi quadrupole splitting. It does \emph{not} follow the lattice symmetry - as is usually the case in NMR. Here we find that the magnetic field appears to set the quantization axes of nuclei and electrons involved in the quadrupole interaction, not the lattice. This leads to very unusual angular dependences and new relaxation phenomena. While a theoretical understanding of the latter effects is still missing, we believe that our results prove that the quadrupole interaction can be used to gain unprecedented local insight into these fascinating materials.

\section{Methods}

\subsection*{Synthesis and characterization}
Single crystals of \ce{Bi2Se3} (S1, S2, S4) were grown using the self-flux method. Elemental Se (99.999\%, \emph{Sigma Aldrich}) and Bi (99.999\%, \emph{Sigma Aldrich}) in three different molar ratios (S1 $-$ Bi$_{1.95}$Se$_3$; S2 $-$ Bi$_{2.00}$Se$_3$; S4 $-$ Bi$_{2.05}$Se$_3$) were filled into quartz tubes under Ar atmosphere. The ampoules were then evacuated and sealed before they underwent heat treatment. The heating scheme reads as follows: melting and reaction at 800$^{\circ}$C for 10 hours followed by a slow cooling and crystal growth period of 100 hours at a cooling rate of 2.5 K/hour, finished by liquid-nitrogen quenching. Large single crystals with a few millimeters in size can easily be cleaved off the ingot. Note that the molar balance between Bi and Se as given in the chemical formulas above denotes the initial molar composition of the melt from which the crystals were grown, rather than the stoichiometry of the final product.
Single crystal S3 (with dimensions $4\times3\times1\mathrm{mm}^3$) is the same as reported in \cite{Georgieva2016} (details about the synthesis can be found there, as well). Microscopic powder of \ce{Bi2Se3} was obtained from grinding commercial \ce{Bi2Se3} flakes (\emph{Sigma Aldrich}) using mortar and pestle.

We performed X-ray diffraction (XRD) of powders from each sample to confirm phase purity. Lattice constants are found to be stable across the samples with $a=\SI{4.142(1)}{\AA}$ and $c=\SI{28.640(4)}{\AA}$. In order to determine residual carrier concentrations in these small band-gap systems, the temperature dependent Hall effect was measured of pieces obtained from pressed and sintered powders as well as of single crystals using a Physical Properties Measurement System (PPMS) from \emph{Quantum Design Inc., USA}. The carrier concentration $n$ was found to be widely temperature independent with values between $n=6.7(5)\times10^{18}\SI{}{cm^{-3}}$ for the Se-rich material (S1) and  $n~=~1.7(1)\times10^{19}\SI{}{cm^{-3}}$ for the Se-deficiency sample (S4).

\subsection*{Experimental techniques}

Magnetic fields for the NMR experiments ranged from 9.4 to 11.74 T, and commercial consoles (Bruker, Tecmag) were  used for excitation and detection. The home-built probes were equipped with single axis goniometers for sample rotation with an accuracy of about $\pm\ang{2}$. Single crystals of millimeter dimension were arranged within the NMR coils inside the goniometer with filling factors close to 1 to ensure maximum signal to noise. The effect on the resonance circuit's quality factor, $Q$, by the insertion of a single crystal into the NMR coil is significant, yielding typical $Q$s of 20 to 30. It is well established that the material shows residual bulk conductivity, but the skin effect can nearly be neglected \cite{Georgieva2016}.

High power solid echo sequences ($\pi/2-\tau-\pi/2$) with typical pulse lengths of $0.75\ \mu s$ were used to record the complete quadrupolar split spectra. The circuit's low quality factor ensures sufficient bandwidths for excitation and detection. Spin echo pulse sequences ($\pi/2-\tau-\pi$) with a $\pi/2$-pulse duration of $10\ \mu s$ were employed for the selective excitation of individual transitions. Free induction decays (FIDs) were recorded of narrow resonance lines as in case of \ce{^2H2O}. The spin lattice relaxation ($1/T_1$) was measured with inversion or saturation recovery of parts of the spectrum (see discussion below). Nutation spectroscopy was used to distinguish different transitions in broad spectra. Here, the pulse power levels were changed rather than the pulse widths in order to keep excitation widths constant.

\subsection*{First Principle Calculations}
Electric field gradients were obtained from electronic structure calculation within density functional theory (DFT) -- using the all-electron full-potential WIEN2k code \cite{wien2k}. The sizes of atomic spheres were chosen as 2.50 and 2.42~a.u.\ for Bi and Se, respectively. All relevant parameters, especially the number of k-points ($18\times18\times18$ mesh) and the size of the basis set ($RK_\mathrm{max}$=14), were properly converged with respect to the value of the electric field gradient of Bi. The spin-orbit interaction was included as second variational method using the scalar-relativistic approximation \cite{MacDonald1980}. The exchange-correlation term was approximated by Perdew-Burke-Ernzenhof variant \cite{Blaha1996} of the generalized gradient approximation (PBE-GGA).


\section{Results}

\subsection{Magnetic field parallel to the crystal $c$-axis and powder spectra}
Given the single Bi site in the unit cell and its local symmetry, cf.~Fig.~\ref{fig:water}B, we expect a single NMR line with a symmetric electric field gradient (EFG) that has its largest principle component ($V_{ZZ}$) along the crystal $c$-axis ($c\parallel Z$), thus $V_{XX}=V_{YY}=-V_{ZZ}/2$. 

\begin{figure}[t]
\centering
\includegraphics[scale=1]{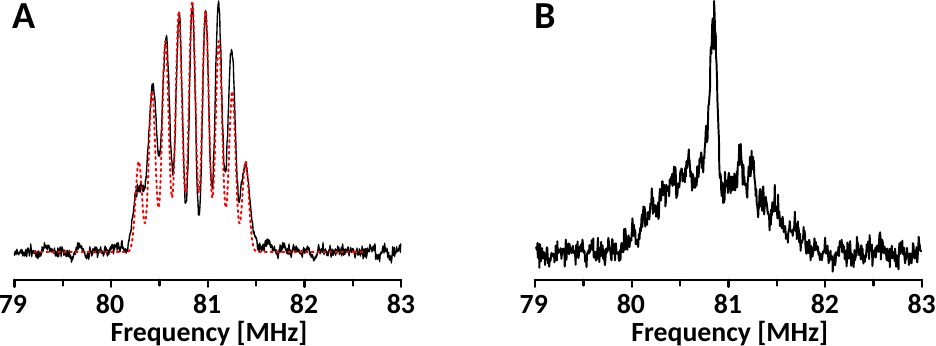}
\caption{\label{fig:BiSpectrum} $^{209}$Bi NMR spectra at \SI{11.74}{T} from Fourier transform of a solid echo with $\pi/2$-pulse widths of \SI{0.75}{\mu s} and a pulse separation $\tau=\SI{15.5}{\mu s}$, obtained from, A, the \ce{Bi2Se3} single crystal S3 at \cpara (the red line is a fit to a $I=9/2$ first-order quadrupole pattern with a magnetic broadening of about 70 kHz) and, B, a powder.}
\end{figure}

Then, if the magnetic field is along the crystal $c$-axis one expects the $^{209}$Bi NMR resonance to be split into $2I = 9$ lines, i.e., a central transition with 4 pairs of satellite lines. 
This is indeed the case, as shown in Fig.~\ref{fig:BiSpectrum}A. We measure a quadrupole splitting of \SI{141(3)}{kHz} between the neighboring transitions (which is found to be slightly sample dependent, see below). Since the splitting is much smaller than the Larmor frequency even at the lowest employed field of \SI{9.4}{T} (i.e. $\sim \SI{64}{MHz}$), the quadrupole interaction needs to be considered only in leading order.

In this very good approximation, the quadrupole Hamiltonian is typically written as,
\beq
\mathcal{H_Q} = \frac{{3I_z^2 - I(I + 1)}}{{4I(2I - 1)}}eQ \cdot {V_{ZZ}}\left\{ {\frac{{3{{\cos }^2}\beta  - 1}}{2} + \frac{\eta }{2}{{\sin }^2}\beta \cos 2\alpha } \right\},
\label{eq:QuadHam}\eeq
if one allows the laboratory frame ($x, y, z$) to be different from that of the principle axes system  ($X, Y, Z$) of the EFG tensor. As usual, $\beta$ is the polar angle between the $z$ and $Z$ axes, $\alpha$ is the azimuthal Euler angle, and $\eta = \left(V_{XX}-V_{YY}\right)/V_{ZZ}$ is the asymmetry parameter. The nuclear spin is denoted by $I$, and the nuclear quadrupole moment by $eQ$. 
We adopt the usual definition of the quadrupole frequency,
\beq
{\nu_Q \equiv \omega _Q/2\pi} = \frac{{3eQ{V_{ZZ}}}}{{2I\left( {2I - 1} \right)h}} \equiv \frac{{3{e^2}qQ}}{{2I\left( {2I - 1} \right)h}},
\label{eq:QuadFreq}
\eeq
and we measure from Fig.~\ref{fig:BiSpectrum}A, with $\beta = 0$ that $\nu_Q \approx \SI{141(3)}{MHz}$. The frequency difference from the central transition for each satellite is given by,
\beq
\Delta \nu_{\rm \sigma} = (\sigma+1/2) \tilde{\nu}_Q(\beta ,\alpha ),
\label{eq:SatFreq}
\eeq
where $\sigma=-1/2$ denotes the central transition ($\sigma = -9/2, -7/2, ..., +7/2$), and $\tilde{\nu}_Q$ the angular dependent quadrupole frequency,
\beq
\tilde{\nu}_Q(\beta ,\alpha ) = {\nu _Q}\left\{ {\frac{{3{{\cos }^2}\beta  - 1}}{2} + \frac{\eta }{2}{{\sin }^2}\beta \cos 2\alpha } \right\}.
\label{eq:QuadRot}
\eeq

Note that a spatial variation of the quadrupole frequency results in a rather typical set of linewidths of the 9 resonances: the central transition would have no broadening, but the outermost satellites would see an amplified broadening by a factor of 4 compared to the innermost satellites. Here, we find the central transition nearly as broad as the outermost satellite transitions, cf.~Fig.~\ref{fig:BiSpectrum}A. We conclude that the principle components of the (traceless) EFG are rather well defined, and we estimate the variation of $V_{ZZ}$ to be less than 5\%, indicating a very homogeneous crystal. This is indeed a surprising finding since for NMR in systems where quadrupole interaction dominates, quadrupole broadened spectra are the rule. In particular for \ce{Bi2Se3} a material known to show considerable deviations from the average stoichiometry, e.g. as evidenced by self-doping \cite{Huang2012}, one would expect quadrupolar broadened spectra, which is clearly not the case.

Also shown, in Fig.~\ref{fig:BiSpectrum}B, is a typical powder spectrum. One sees immediately -- we used the same frequency scale as for the single crystal spectra in Fig.~\ref{fig:BiSpectrum}A -- that the powder spectrum has a similar width compared to the single crystal. This is \emph{not} expected since the angular dependence in a powder does not favour crystallites with directions $\beta \approx 0$ (from near poles of the unit sphere), rather those near the equator ($\beta \approx \SI{90}{^o}$) determine the spectrum. So we expect for the powder spectrum roughly half the width in view of \eqref{eq:QuadRot}. This fact points to an unusual angular dependence, already.
\begin{figure}
\centering
\includegraphics[width=1\textwidth]{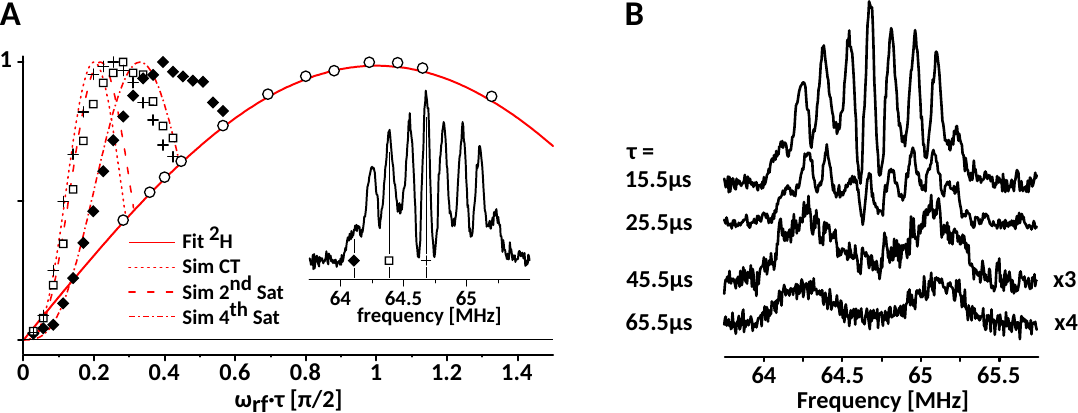}
\caption{\label{fig:SEdecay_nutC} A, Selective spin echo nutation measurements with $\pi/2$-pulse length of \SI{10}{\mu s} shown for the central transition ($+$), the $2^{\mathrm{nd}}$ (open squares),  and the $4^{\mathrm{th}}$ (solid diamonds) lower satellite (also done for all the other transitions), and a rescaled nonselective $^{2}$H nutation in \ce{D2O} (open circles). The deuterium nutation was fitted with a $\sin(x)$ function (red solid line) and from this result the nutations of selectively excited transitions of a quadrupolar split spin 9/2 were simulated (red dashed lines). Intensities were normalized for the sake of clarity. B, the $^{209}$Bi spectrum (Fig.\ref{fig:BiSpectrum}, a) obtained from solid echo measurements with increasing pulse separation $\tau$ as indicated. The lower two spectra (\SI{45.5}{\mu s} and \SI{65.5}{\mu s}) have been magnified by a factor of 3 and 4, respectively.
}
\end{figure}
\begin{figure}
\centering
\includegraphics[width=0.8\textwidth]{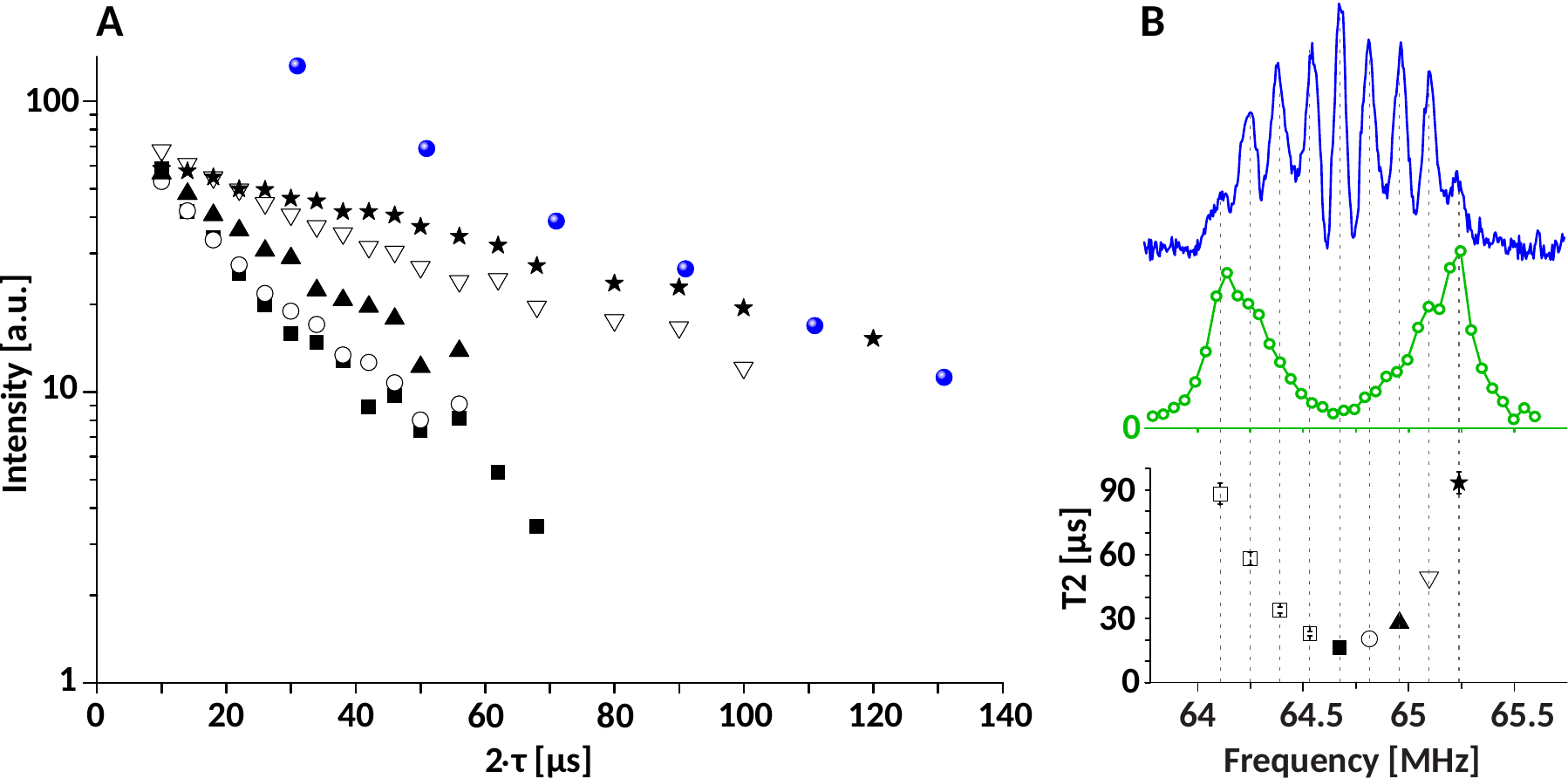}
\caption{\label{fig:T2} A, spin echo decays of selectively excited transitions ($\pi/2$-pulse lengths of \SI{10}{\mu s}) represented by the black symbols (for assignment of the symbols cf. bottom panel in B). The blue data points represent the full spectral intensity of the solid echo decay in Fig.~\ref{fig:SEdecay_nutC}B. Upper panel, B: nonselective solid echo measurement (blue) with $\tau=\SI{15.5}{\mu s}$ and the intensity pattern of frequency stepped selective spin echoes (green data points) with $\pi/2$-pulse lengths of \SI{10}{\mu s} and $\tau=\SI{45}{\mu s}$; lower panel, B: $T_2$ values obtained from fittings to the spin echo decays shown in A, obtained from simple exponential functions.
}
\end{figure}

To be certain that the observed resonances are due to first-order quadrupole effects, we performed nutation experiments on the individual $^{209}$Bi lines, and compared the results with those obtained for $^2$H NMR of deuterated water (the $^2$H resonance is about \SI{2.9}{MHz} below that of $^{209}$Bi at \SI{9.4}{T} and can thus serve as a reference -- after correction for $Q$ -- for the non-selective radio frequency amplitude that is difficult to measure for the broad Bi spectrum). The results are shown in Fig.~\ref{fig:SEdecay_nutC}A. Excellent agreement with the expected effective radio frequency amplitudes for selective excitation is found for not too large power (the pulses for the selective spin echo experiments were \SI{10}{\mu s} and \SI{20}{\mu s} for the $\pi/2$ and $\pi$ pulse, respectively, and the nutation was recorded as a function of the power level). Deviations at larger power levels are expected, given the narrowly spaced resonances and the excitation of the accompanied forbidden transitions \cite{Haase1994}.

In Fig.~\ref{fig:SEdecay_nutC}B we present the changes of the spectrum as a function of the pulse separation for the non-selective echo sequence used in Fig.~\ref{fig:BiSpectrum}. We observe a rather rapid decay with a $T_2 \sim \SI{35}{\mu s}$, accompanied by significant spectral changes. Most notably, the central region of the spectrum decays more rapidly than the outer satellites. This leads to the pronounced dip in the center of the spectra.

In a next experiment we measured the spin echo decay of the selectively excited transitions with a $\pi/2-\tau-\pi$ echo, cf. Fig.~\ref{fig:T2}{{A}}. We observe nearly exponential decays and the selective $T_2$'s vary by about a factor of 5, i.e., the central transition decay is about 5 times as fast as that of the outermost lines. The overall non-selective $T_2$ is not very different, and for small $\tau$ similar  to that of the selectively excited central region. This shows that the decay is rather independent on the number of flipped neighbors.

Also shown, in Fig.~\ref{fig:T2}{{B}}, is a spectrum recorded with frequency stepped selective spin echoes. Clearly, the spectral features are lost and the faster decay in the central region leads to the double-horn spectrum that was reported earlier \cite{Young2012,Nisson2014}.

\begin{figure}
\centering
\includegraphics[width=0.5\textwidth]{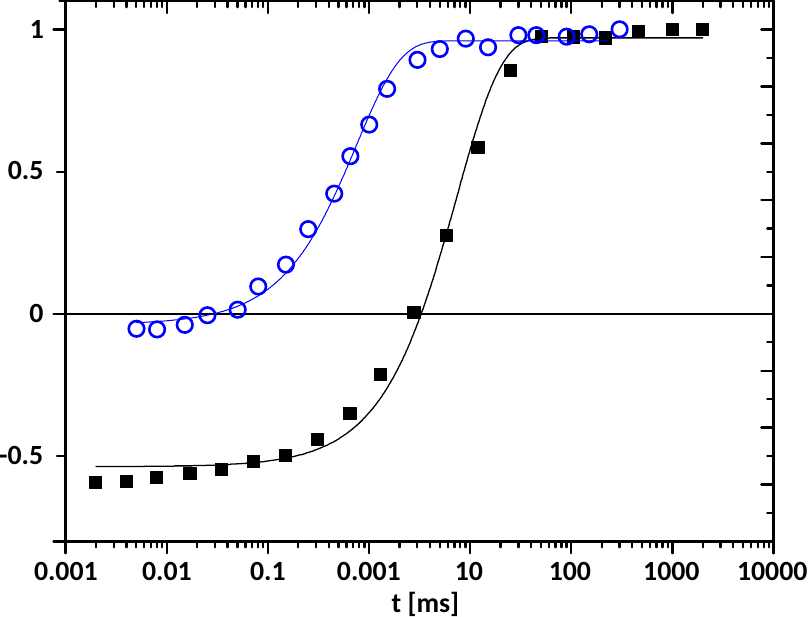}
\caption{\label{fig:T1} Saturation recovery of the selectively excited central transition (open, blue) for $c||B_0$, and inversion recovery (solid, black) near the magic angle for a broader non-selectively excited region, both obtained with sample S3. Solid lines represent simple, single exponential fits to $M(t)=(1-f\exp(-t/T_1))$ resulting in apparent $T_1^{CT}=\SI{0.75(5)}{ms}$ and a non-selective $T_1=\SI{7.5(5)}{ms}$.
}
\end{figure}

Due to the rapid spin echo decays and broad spectral features, the precise measurement of nuclear relaxation is difficult. In Fig.~\ref{fig:T1} we show the results of a selectively measured saturation recovery for \cpara, as well as a nearly non-selective inversion recovery near the magic angle for a single crystal (S3). One expects that the recovery after a nearly full saturation (or inversion) is considerably slower for a spin 9/2 nucleus (since all spins are out of equilibirum) than the recovery of a selectively saturated (or inverted) central transition. This is what we observe, but we cannot exclude spectral diffusion for the selectively excited transitions. Also, we cannot say whether the observed, fast relaxation is magnetic or quadrupolar in origin.

\begin{figure}
\centering
\includegraphics[width=.8\textwidth]{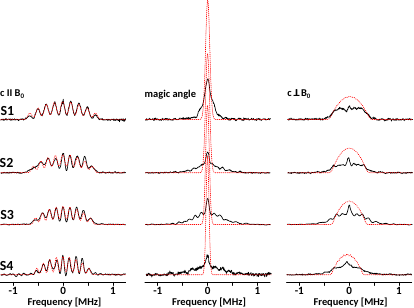}
\caption{\label{fig:ADcrystals}  Angular dependent spectra (black) obtained from nonselective solid echoes ($\pi/2$-pulse length \SI{0.75}{\mu s}) for the four single crystals (S1, S2, S3, S4) at three crystal orientations: for \cpara (left), the magic angle (middle) and for \cperp (right). Carrier concentration increases from top to bottom. Magnetic shifts have been subtracted. Relative intensities as measured. Red dashed lines represent first-order quadrupole patterns fitted at \cpara and their corresponding axial symmetric angular dependence at the magic angle and \cperp
}
\end{figure}

We turn to the investigation of the angular dependences of the spectra, now.

\subsection{Angular dependences of single crystal spectra}
Spectra for the four different samples (S1, S2, S3, S4) are shown in Fig.~\ref{fig:ADcrystals}, recorded at three different angles $\beta$: first, for $\beta = 0$, i.e. with the field parallel to the crystal $c$-axis (\cpara); second, near the magic angle ($\beta \approx \ang{54.7}$); and  third, for $\beta = \SI{90}{^o}$, i.e. with the field in the plane ($\cperp$). While the spectra for \cpara agree with the expected behavior, the spectra near the magic angle show the largest inconsistencies.

Sample S1, while similar to what one expects from \eqref{eq:QuadRot}, shows unexpected broadening at the magic angle and a missing intensity of about 30\% (about 35\% of the intensity are missing for \cperp).

For sample S2 the deviations from the expected behavior are more obvious. There is substantial intensity in form of a broad resonance even at the magic angle (of up to $\pm\SI{500}{kHz}$ away from the center). A somewhat larger amount of intensity is missing. Note also that there is an increased discrepancy for \cperp, as the observed width is much larger than what is expected (about 1/2 of that for \cpara). Similar observations are made for samples S3 and S4, i.e., while the spectra for \cpara appear to fit the symmetric quadrupole pattern, the lineshapes at the magic angle, as well as for \cperp are in stark disagreement with such an explanation (changes in $T_2$ cannot explain the discrepancies).

\begin{figure}
\centering
\includegraphics[width=.4\textwidth]{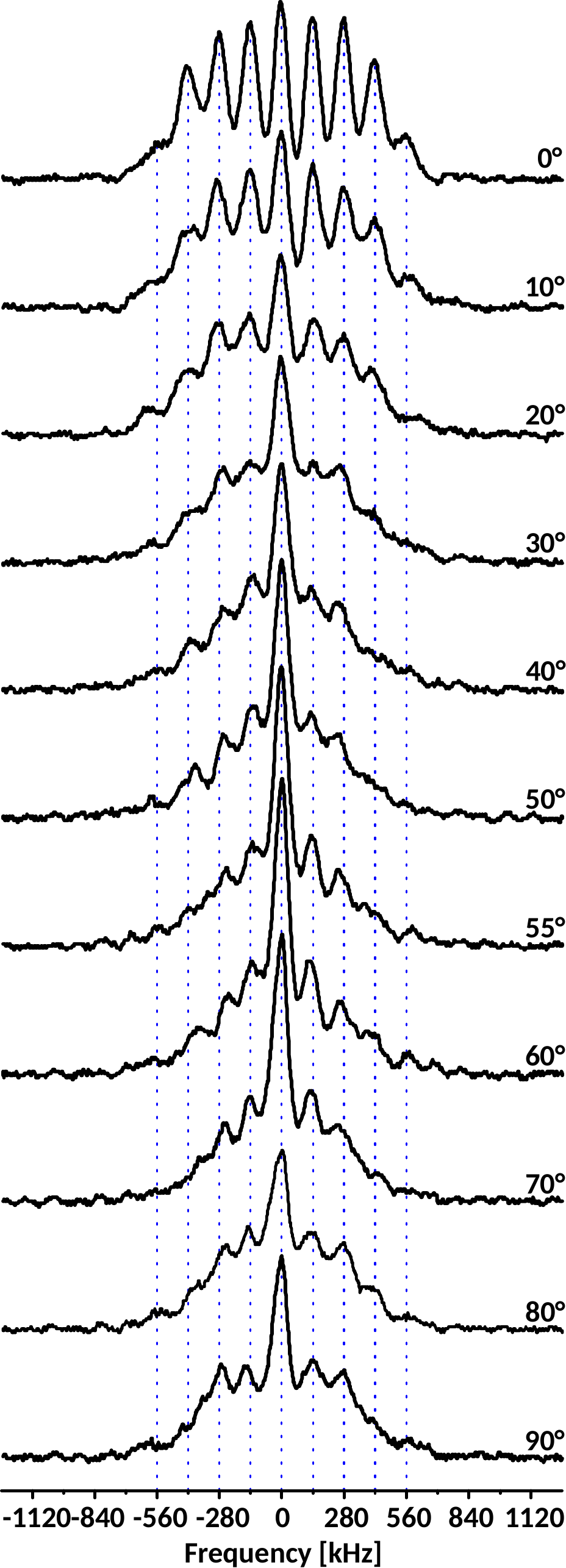}
\caption{\label{fig:AD}Angular dependent spectra of single crystal (S3) using the solid echo sequence with pulse length and separation of \SI{0.75}{\mu s} and \SI{15.5}{\mu s}, respectively. \ang{0} corresponds to \cpara, \ang{90} represents the \cperp orientation. The magnetic shift has been subtracted for each measurement. Measurements taken at \SI{9.39}{T}.
}
\end{figure}

A more detailed angular dependence for sample S3 is presented in Fig.~\ref{fig:AD}. Here it becomes obvious that the splittings as a function of $\beta$ do not agree with \eqref{eq:QuadRot}. Rather, one is inclined to conclude, based on the positions of peaks and valleys, that the broad intensity - irrespective of the angle of rotation - represents the same satellite transitions. This, however, would be in striking disagreement with what one expects from angular dependent quadrupole spectra.

\begin{figure}
\centering
\includegraphics[width=.5\textwidth]{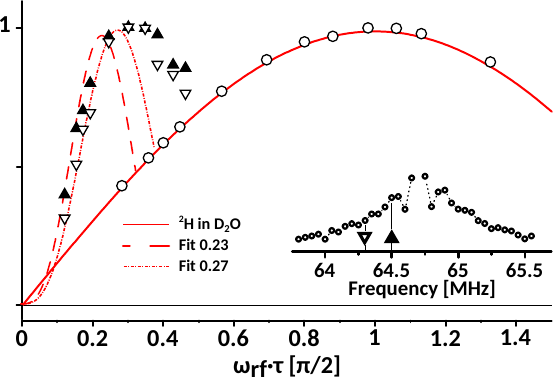}
\caption{\label{fig:NutMagic} Nutation measurements obtained from single crystal (S3) at the magic angle using selective spin echoes with $\pi/2$-pulse length of \SI{10}{\mu s}. Open circles and the red solid line reproduce the $^{2}$H nutation in \ce{D2O} from Fig.\ref{fig:SEdecay_nutC}, (A). Triangles represent the position of the nutation measurement in frequency units as indicated in the inset. They correspond to \SI{200}{kHz} and \SI{400}{kHz} away from the center of the complete spectrum (located at \SI{64.7}{MHz}). The data sets have been fitted (red dashed lines) with emphasis on small power levels resulting in maxima at 0.23 and 0.27 clearly representing a quadrupolar split system even at the magic angle. The inset gives frequency swept and $T_2$-corrected spin echo intensities with $\tau=\SI{10}{\mu s}$ at the magic angle.
}
\end{figure}

\begin{figure}
\centering
\includegraphics[width=0.4\textwidth]{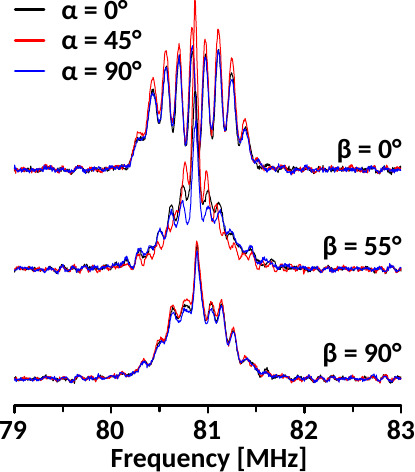}
\caption{\label{fig:Alpha}Angular dependent spectra of single (S3) using nonselective solid echoes. For each orientation $\beta$ spectra have been taken for three different single crystal orientations within the coil, rotated about its respective crystal $c$-axis. This corresponds to a change of $\alpha$ in eq. \eqref{eq:QuadRot} as indicated in the legend and thus proves a symmetric EFG. Here, $\alpha$ has no direct correspondence with the crystal $a$ or $b$-axes. Measurements taken at \SI{11.74}{T}.
}
\end{figure}

In order to learn more about these broad resonances, we performed NMR nutation experiments -- similar to what is presented in Fig.~\ref{fig:SEdecay_nutC}, but now recorded near the magic angle (as far away in frequency as  $\pm\SI{600}{kHz}$ from the center). Selected results are shown in Fig.~\ref{fig:NutMagic}. 
Indeed, the intensity outside the central region appears to belong to satellite transitions, similar to what one finds for \cpara.  Note that given the lattice symmetry, one demands that the asymmetry parameter is vanishingly small ($\eta \approx 0$) so that the spectra must narrow considerably near the magic angle. 
If this is true ($\eta \approx 0$), a rotation of the sample about the crystal $c$-axis for different $\beta$ should leave the spectra largely unchanged (a change of $\alpha$ in \eqref{eq:SatFreq}).
The results are shown in Fig.~\ref{fig:Alpha}. Since there are only minor changes to the spectra (some of them cannot be avoided since we change the orientation of the sample in the coil), we conclude on the rotational symmetry of the in-plane EFG components, i.e. we have indeed a nearly symmetric tensor independent on the orientation of the crystal with respect to the field.

The above experiments showed that the usual understanding of the spectra in terms of a quadrupole Hamiltonian \eqref{eq:QuadHam} cannot explain the data, i.e., both coordinate systems ($x, y, z$ and $X, Y, Z$) do not seem to rotate with respect to each other by changing the direction of the magnetic field with respect to the crystal axes.


\section{Discussion}
We begin the discussion with the spectrum for \cpara in Fig.~\ref{fig:BiSpectrum}. As mentioned above, in reference to the chemical structure we expect such a quadrupolar split set of lines. We find a systematic change of the quadrupole splitting from \SI{164(2)}{kHz} to \SI{128(3)}{kHz}, while the crystal lattice parameters obtained from XRD do not change across the different samples. And since samples S1, S2, and S4 were prepared with different molar ratios of Bi and Se atoms in order to drive self-doping effects, yielding samples with increasing excess carrier concentration $n$, we conclude that the EFG at $^{209}$Bi in \ce{Bi2Se3} is related to $n$, as well. 

These two facts and the absence of a significant inhomogeneity, we believe, point to the special band structure of topological insulators. It is known that the energy band structure near the ${\Gamma}$-point in the Brillouin zone is inverted due to spin-orbit coupling \cite{Zhang2009}. As a direct consequence for \ce{Bi2Se3}, Bi and Se states are being exchanged to some extent, which must reduce the EFG at the Bi site. Furthermore, extra electrons brought into the system by self-doping do inevitably emerge on the energetically lower most edges of the conduction band, namely in the vicinity of the $\Gamma$-point \cite{Zhang2009}. Thus, free carriers in this system must change the quadrupole splitting while reflecting the real-space characteristics of the energy band inversion.

\begin{figure}
\centering
\includegraphics[width=0.75\textwidth]{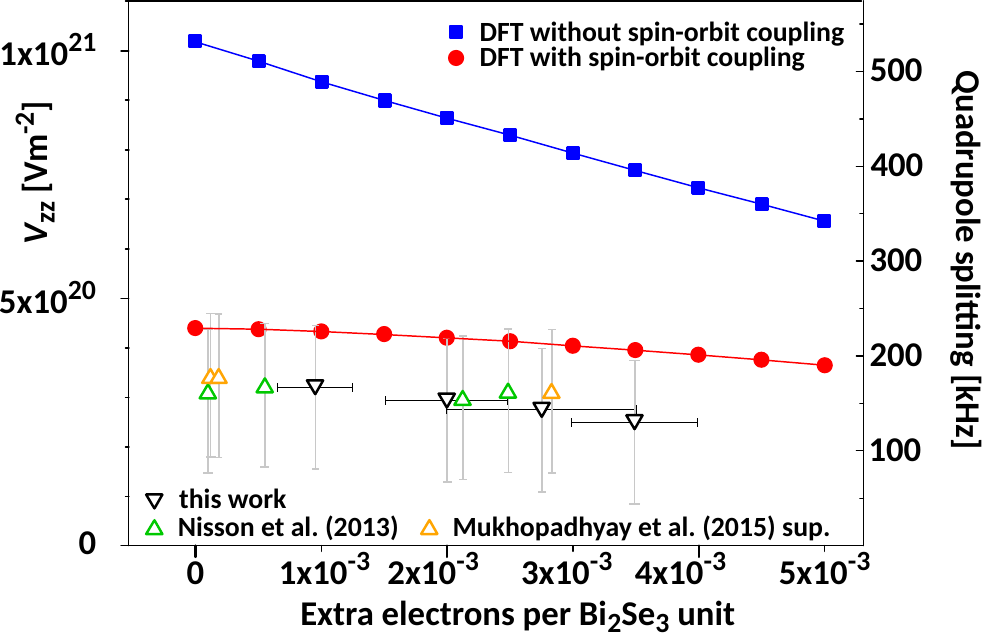}
\caption{\label{fig:DFT}The local EFG at the Bi crystal site ($V_{ZZ}$) as obtained from DFT calculations with (red) and without (blue) spin-orbit coupling (SOC). Colored triangles give $V_{ZZ}$ as obtained from measured quadrupole splittings (including literature data \cite{Nisson2013}, and  supplementary of \cite{Mukhopadhyay2015}) using eq.~\eqref{eq:QuadFreq} and the quadrupole moment of $|Q_{\mathrm{Bi}}|=\SI{516}{mb}$ \cite{Bieron2001}. The grey error bars indicate the variation of $V_{ZZ}$ when assuming other reported $|Q_{\mathrm{Bi}}|$ ranging from \SI{370}{mb} to \SI{710}{mb} \cite{Bieron2001}. The black data points display the results of this work according to Hall-effect and shift measurements (Fig. S 6 of \cite{Mukhopadhyay2015}, supplementary information).
} 
\end{figure}

In order to investigate this quantitatively, first-principle calculations with the Wien2k code \cite{wien2k} were performed. Some results are summarized in Fig.~\ref{fig:DFT} and show, indeed, the role of spin-orbit coupling on the EFG. 
The carrier concentrations were modeled by adding small amounts of electronic charge into the rhombohedral \ce{Bi2Se3} unit cell: extra charges up to 0.005$e$ cover the range of carrier densities in the studied samples. In order to preserve the neutrality of the unit cell, this additional charge was compensated by the same amount of positive background charge. 
Without spin-orbit interaction the extra charge visibly decreases the calculated Bi EFG: the extra charge occupies predominantly the $6p_z$-states of Bi and the $p$-$p$ term is the dominant contribution to the EFG for Bi. Since the Bi $p_z$-states are initially less occupied than the $p_x$- and $p_y$-states, the $p$-anisotropy decreases and we observe a monotonous decrease of the EFG tensor on Bi.

When the spin-orbit interaction is enabled, band inversion occurs -- accompanied by charge transfer from $p$-states of Se to Bi. The occupation of Se $p_z$-states for both crystal sites is considerably reduced while the $6p_z$-states of Bi are now more populated. As a consequence, the EFG on Bi abruptly decreases. The extra charge is then added also to $p_x$- and $p_y$-states, which leads to a less pronounced decrease of the EFG. 

This explains our experimental results (note, that in the experiments we can only determine the absolute value of $V_{{ZZ}}$ through $\nu_{\mathrm{Q}}$, not its sign).
The calculated EFG is slightly larger than the experimentally observed one. Most likely, this discrepancy is  due to imperfect optimization of the crystal lattice because of the presence of van der Waals bonds (GGA neglects dispersion interactions, only the volume was optimized, while the $c/a$ ratio in hexagonal unit cell representation was kept fixed at value from the experiment). We estimate other sources of errors to be of the order of 0.01 (in units of $10^{21}$~Vm$^{-2}$).
We conclude that the $^{209}$Bi NMR quadruple spectrum for \cpara is well understood in terms of the electronic structure.

The rapid changes in the lineshapes with pulse separation (spin echo decay) were observed in earlier publications \cite{Young2012,Nisson2014}, but not fully acknowledged (\cite{Nisson2014} assumed a Redfield term). The large spectral width from the quadrupole interaction together with the rapid $T_2$ demand a high time resolution for NMR measurements, which is not readily available or limits signal to noise. In our experiments, by resorting to small sample volumes of $5-\SI{10}{mm^3}$, we could separate the rapid $T_2$ from the changes in the lineshapes due to the angular dependence, sufficiently. 
On a qualitative level, we believe that the recently discovered large indirect spin-spin coupling between nuclei \cite{Georgieva2016} is responsible for the rapid $T_2$ of Bi, in particular that it affects the central levels of the spectrum much more than the outer ones (a factor of almost 5 in Fig.~\ref{fig:T2}). Such behavior is expected if the quadrupole interaction is not large compared to the indirect coupling, so that spin flips in the central levels are not fully suppressed \cite{Haase1993}.  

We now turn to the conundrum of the angular dependence of the $^{209}$Bi NMR spectra. Most surprising is the observation that the quadrupole pattern observed for \cpara does not collapse near the magic angle, which is most obvious for the crystals with higher carrier concentrations ($n > 10^{19} {\rm cm^{-3}}$). We find that this effect is in agreement with hitherto published data \cite{Young2012, Nisson2013, Nisson2014}, however, the problem was not addressed in detail, before. 

One may seek an explanation for the unusual behavior in local imperfections that often cause quadrupolar broadening. In fact, the coordination of Se neighbors from a Bi atom point of view encloses an angle of $\ang{50}-\ang{60}$ with the crystal $c$-axis, i.e. close to the magic angle. Thus, the inhomogeneities arising from Se vacancies would not be very effective for \cpara (when investigating Bi nuclei). However, one key problem with such attempts is that due to the measured intensities, the inhomogeneities must affect large numbers of Bi atoms, which appears highly unlikely since each Se vacancy adds 2 electrons, hence, from Fig.\ref{fig:DFT}, only about 1 out of 1000 Se atoms is missing. Only extended charge density waves scenarios could affect more nuclei, but these would certainly cause quadrupole broadened spectra also for \cpara, which are not observed. In addition, the various samples show very similar high quality patterns for \cpara, but not at other angles. 
Finally, the broad signal -- even at the magic angle -- appears to be single crystal like, similar to what one has for \cpara.
The detailed angular dependence in Fig.~\ref{fig:AD} suggests that the spectra do not change very much at all if one turns the sample with respect to the magnetic field. 

The fact that a random powder spectrum, cf.~Fig.~\ref{fig:BiSpectrum}B, has a much larger width than what follows from the \cpara spectra clearly prohibits an explanation in terms of a regular orientational dependence, already.

One might then be tempted to assume that the EFG consists of two terms. One term is caused by the chemical structure, e.g. from neighboring ionic charges, and shows the usual angular dependence given by \eqref{eq:QuadRot}. This background splitting, one would assume, should follow in the limit of zero carrier concentration, cf.~Fig.~\ref{fig:DFT}. Then, the second part would be due to itinerant carriers and has a negative sign. For \cpara the splitting is given by the sum of the two terms thus one could explain the observed shape. As one rotates the sample the second term should lead to the deviation from the regular angular dependence. However, we were not able to fit the data with such a scenario. The difficulty arises from the fact that the background term is large to begin with, and it should always disappear at the magic angle. This is not what is observed. Rather as the component with an unusual angular dependence grows, it also appears to affect the background EFG. 
These findings point to a single, dominant process from electrons near the Fermi surface also for the first term and there is essentially \emph{no} lattice EFG (in agreement with the nearly cubic coordination of the six nearest Se atoms forming an octahedron, or the fact that we do not observe quadrupolar broadening in these inhomogeneous materials).\par\medskip

Therefore, all these observations led us to discard the usual picture and propose a very different scenario: an axially symmetric EFG tensor that follows the external field $B_0$ rather than the crystal $c$-axis. A possible line of (qualitative) arguments for this scenario is presented now.

The electric quadrupole interaction involves the nuclear spin. Therefore, it is written conveniently in terms of spherical tensors \cite{Sobelman2012}, i.e., 
\beq
\mathcal{H_Q} = \sum\limits_{q =  - 2}^{ + 2} {{{\left( { - 1} \right)}^q}{Q^{(2)}_{- q}}{V^{(2)}_{q}}}.
\eeq
In this notation, one defines the nuclear quadrupole moment $eQ$ by, 
\beq
eQ = 2\bra{I\,m_I = I}{Q_{0}^{(2)}}\ket{I\,m_I=I},
\eeq
where \ket{I m_I = I} is the eigenket of the nuclear spin ($I$) with the largest magnetic quantum number $m = I$.
Similarly, the electric field gradient, $eq$, created by an electron in the state $\ket{J, m_J}$ is defined by,
\beq
eq =2\left\langle {J\,{m_J} = J} \right|{V^{(2)}_{0}}\left| {J\,{m_J} = J} \right\rangle,
\eeq
where $J$ is its total angular momentum of the electron.
In terms of the spherical electronic coordinates ($r_e, \theta_e, \phi_e$) one has,
\beq
V^{(2)}_0 =\sqrt{\frac{4\pi}{5}}\int{\frac{\rho_e}{r_e^3}Y_{20}(\theta_e,\phi_e)d\tau_e}.
\eeq
In a solid, one imagines an effective electronic wave function, \ket{\psi_e}, since the state of the electrons is not known precisely, and one writes,
\beq
\left\langle {I,{m_I},{\psi _e}} \right|{H_Q}\left| {I,{m_I},{\psi _e}} \right\rangle  = \sum\limits_{q =  - 2}^{ + 2} {{{\left( { - 1} \right)}^q}\left\langle {I,{m_I}} \right|{Q^{(2)}_{- q}}\left| {I,{m_I}} \right\rangle \left\langle {{\psi _e}} \right|{V^{(2)}_{q}}\left| {{\psi _e}} \right\rangle }.
\eeq
Only the term with $q=0$ contributes \cite{Sobelman2012}, i.e.,
\beq
\left\langle {I,{m_I},{\psi _e}} \right|{H_Q}\left| {I,{m_I},{\psi _e}} \right\rangle =  \frac{{3{m^2} - I(I + 1)}}{{2I(2I - 1)}}eQ \cdot \left\langle {{\psi _e}} \right|{V^{(2)}_{0}}\left| {{\psi _e}} \right\rangle.
\eeq
In a strong magnetic field $B_0$ in $z$-direction the nuclear spin is quantized along the field axis ($z$-axis), and we can write,
\beq
\mathcal{H_Q}_{\rm eff} =  \frac{{3{I_z^2} - I(I + 1)}}{{2I(2I - 1)}}eQ \cdot \left\langle {{\psi _e}} \right|{V^{(2)}_{0}}\left| {{\psi _e}} \right\rangle.
\eeq
In systems without significant spin-orbit coupling one can neglect the electronic spin (that does align with the field) and consider only the resulting electronic orbital angular momentum. Then, for a given shell, it can be factorized into a radial and an angular part. The latter is tied to the chemical bonding and thus typically fixed in the crystal's unit cell. This makes the quadrupole coupling dependent on the local crystal symmetry, and yields the typical treatment of quadrupole coupling in solids, cf.~\eqref{eq:QuadHam}.

We argue that in strongly spin-orbit coupled systems the situation might be different. Here, the electron has the total electron angular momentum $\bf{J}$ as a good quantum number, and it may follow the field, loosely speaking. This must have consequences also for the electric field gradient $eq =2\left\langle {J\,{m_J} = J} \right|{V^{(2)}_{0}}\left| {J\,{m_J} = J} \right\rangle$, and such electrons should contribute to the EFG differently (in a simple molecule one can use the Wigner-Eckart theorem to relate the actual EFG to other quantum numbers). In particular, such an EFG should depend on the field axis, as well.   

We think that this kind of behavior must be behind the observations for the angular dependences shown above. \ce{Bi2Se3} has certainly strong spin-orbit coupling with a $g$-factor of about 32 for \cpara and 23 for \cperp \cite{Kohler1975}. 
While our explanation is lacking quantitative theory, we believe that the quadrupole interaction in these spin-orbit coupled systems must be fundamentally different. To our knowledge, this has never been proposed. 
Unfortunately, due to the design of the DFT calculations that limits the spin-orbit interaction to be accounted for only within the atomic spheres, we cannot model these electrons with large $g$-factors, or an angular dependence.

Corroborating evidence, we believe, is given by a recent report on electron paramagnetic resonance (EPR) in \ce{Bi2Se3} that shows spin-1/2 resonances with similar $g$--factors \cite{Wolos2016}. In a sense this shows the Larmor precession of a large $g$--factor electron (with consequences for the quadrupole coupling).

Finally, if our conjecture is true, a finite life time of the involved electronic states should lead to a new relaxation mechanism in particular for quadrupolar nuclei, in analogy to (magnetic) spin-rotation interaction. Since the quadrupole interaction exceeds the magnetic coupling, the nuclear relaxation of Bi should then be fast and quadrupolar. Indeed, while the nuclear relaxation of Se spin-1/2 nuclei in \ce{Bi2Se3} is rather long (various seconds) \cite{Georgieva2016}, the Bi relaxation is rather short, cf.~Fig.~\ref{fig:T1}, and it was reported to show unusual temperature dependences, earlier \cite{Young2012,Nisson2013}. Typically, quadrupolar relaxation is assumed to be due to phonons, but one needs Raman processes to account (qualitatively) for large rates that do not follow from single phonon scattering. The process suggested here is expected to be more effective since it couples directly to the electronic density of states, and a comparably long life time of the involved electrons can be much closer to the Larmor precession rate than in the case of phonons, and thus be more effective.

\section{Conclusion}
We investigated $^{209}$Bi NMR quadrupole splittings for a number of \ce{Bi2Se3} samples with different carrier concentrations. If the magnetic field $B_0$ is parallel to the crystal $c$-axis (\cpara) we observe 9 equally broadened resonance lines as for an axially symmetric electric field gradient (EFG) with its largest $Z$-axis parallel to $c$ for this $I=9/2$ nucleus. Despite an expected chemical inhomogeneity the lines are only magnetically broadened. The quadrupole splittings decrease slightly with increasing carrier concentration. The data are in quantitative agreement with first-principle calculations when the magnetic field is along the crystal $c$-axis (for which calculations are possible), and comparison shows that the band inversion (near the $\Gamma$-point) due to the spin-orbit coupling decreases the quadrupole splitting as the inversion changes the occupation between Bi and Se orbitals. 

As already the comparison of powder spectra with single crystal data for \cpara shows, we discovered a non-trivial angular dependence of the quadrupole splitting if the $B_0$ is rotated with respect to $c$. While the spectra appear to maintain axial symmetry around $c$, the EFG's $Z$-axis appears to follow $B_0$, i.e. it does not remain parallel to the crystal $c$-axis, as is typically the case for NMR of quadrupolar nuclei. 

We propose that the special properties of the strongly spin-orbit coupled electrons in these topological materials must be behind this effect, and that the field leads to a non-trivial change of their quantization axis.
Furthermore, since the electronic life time of these electrons leads to a time-dependent EFG, a new source of nuclear relaxation is proposed, which appears to be in agreement with the observations.

While a more detailed quantitative theory is missing, the results show that NMR can provide a quantitative measure of the band inversion, and quite possibly it holds further important information about these new class of materials.

\section*{Acknowledgement}
We acknowledge stimulating discussions with I. Garate (Sherbrooke), U. Zuelicke (Wellington), O. Sushkov (Sydney), B. Fine (Moscow), M. Geilhufe (Stockholm), B. Buckley (Wellington), B. Mallet (Auckland), N. Georgieva, M. Jurkutat, and J. Nachtigal (Leipzig) and for help with XRD M. Ryan (Wellington) and H. Auer (Leipzig). R.G. acknowledges the funding by the MacDiarmid Institute. We thank the Universit\"at Leipzig and the Deutsche Forschungsgemeinschaft for financial support.

\section*{Bibliography}
\bibliography{references}

\begin{thebibliography}{34}
\providecommand{\natexlab}[1]{#1}
\providecommand{\url}[1]{\texttt{#1}}
\expandafter\ifx\csname urlstyle\endcsname\relax
  \providecommand{\doi}[1]{doi: #1}\else
  \providecommand{\doi}{doi: \begingroup \urlstyle{rm}\Url}\fi

\bibitem[Zhang et~al.(2009)Zhang, Liu, Qi, Dai, Fang, and Zhang]{Zhang2009}
Haijun Zhang, Chao-Xing Liu, Xiao-Liang Qi, Xi~Dai, Zhong Fang, and Shou-Cheng
  Zhang.
\newblock {Topological insulators in Bi$_2$Se$_3$, Bi$_2$Te$_3$ and
  Sb$_2$Te$_3$ with a single Dirac cone on the surface}.
\newblock \emph{Nature Physics}, 5\penalty0 (6):\penalty0 438--442, May 2009.

\bibitem[Emprechtinger et~al.(2013)Emprechtinger, Lis, Rolffs, Schilke, Monje,
  Comito, Ceccarelli, Neufeld, and van~der Tak]{Emprechtinger2012}
M.~Emprechtinger, D.~C. Lis, R.~Rolffs, P.~Schilke, R.~R. Monje, C.~Comito,
  C.~Ceccarelli, D.~A. Neufeld, and F.~F.~S. van~der Tak.
\newblock {THE ABUNDANCE, ORTHO/PARA RATIO, AND DEUTERATION OF WATER IN THE
  HIGH-MASS STAR-FORMING REGION NGC 6334 I}.
\newblock \emph{The Astrophysical Journal}, 765\penalty0 (1):\penalty0 61,
  March 2013.

\bibitem[Kilaj et~al.(2018)Kilaj, Gao, R{\"o}sch, Rivero, K{\"u}pper, and
  Willitsch]{Kilaj2018}
Ardita Kilaj, Hong Gao, Daniel R{\"o}sch, Uxia Rivero, Jochen K{\"u}pper, and
  Stefan Willitsch.
\newblock {Observation of different reactivities of para and ortho-water
  towards trapped diazenylium ions}.
\newblock \emph{Nature Commun.}, pages 1--7, May 2018.

\bibitem[Meier et~al.(2018)Meier, Kou{\v{r}}il, Bengs, Kou{\v{r}}ilov{\'a},
  Barker, Elliott, Alom, Whitby, and Levitt]{Meier2018}
Benno Meier, Karel Kou{\v{r}}il, Christian Bengs, Hana Kou{\v{r}}ilov{\'a},
  Timothy~C Barker, Stuart~J Elliott, Shamim Alom, Richard~J Whitby, and
  Malcolm~H Levitt.
\newblock {Spin-Isomer Conversion of Water at Room Temperature and
  Quantum-Rotor-Induced Nuclear Polarization in the Water-Endofullerene
  H$_2$O@C$_{60}$}.
\newblock \emph{Phys. Rev. Lett.}, 120\penalty0 (26):\penalty0 266001, June
  2018.

\bibitem[Lipsicas and Bloom(1961)]{Lipsicas1961}
M.~Lipsicas and M.~Bloom.
\newblock Nuclear magnetic resonance measurements in hydrogen gas.
\newblock \emph{Canadian Journal of Physics}, 39\penalty0 (6):\penalty0
  881--907, 1961.
\newblock \doi{10.1139/p61-095}.
\newblock URL \url{https://doi.org/10.1139/p61-095}.

\bibitem[Abragam(1961)]{Abragam1961}
Anatoly Abragam.
\newblock \emph{Principles of Nuclear Magnetism}.
\newblock Oxford University Press, 1961.

\bibitem[Oppenheim and Bloom(1961)]{Oppenheim1961}
Irwin Oppenheim and Myer Bloom.
\newblock Nuclear spin relaxation in gases and liquids: I. correlation
  functions.
\newblock \emph{Canadian Journal of Physics}, 39\penalty0 (6):\penalty0
  845--869, 1961.
\newblock \doi{10.1139/p61-093}.
\newblock URL \url{https://doi.org/10.1139/p61-093}.

\bibitem[Xia et~al.(2009)Xia, Qian, Hsieh, Wray, Pal, Lin, Bansil, Grauer, Hor,
  Cava, and Hasan]{Xia2009}
Y.~Xia, D.~Qian, D.~Hsieh, L.~Wray, A.~Pal, H.~Lin, A.~Bansil, D.~Grauer, Y.~S.
  Hor, R.~J. Cava, and M.~Z. Hasan.
\newblock {Observation of a large-gap topological-insulator class with a single
  Dirac cone on the surface}.
\newblock \emph{Nature physics}, 5\penalty0 (6):\penalty0 398, 2009.
\newblock URL \url{https://doi.org/10.1038/nphys1274}.

\bibitem[Podorozhkin et~al.(2015)Podorozhkin, Charnaya, Antonenko,
  Mukhamad'yarov, Marchenkov, Naumov, Huang, Weber, and
  Bugaev]{Podorozhkin2015}
D.~Yu. Podorozhkin, E.~V. Charnaya, A.~Antonenko, R.~Mukhamad'yarov, V.~V.
  Marchenkov, S.~V. Naumov, J.~C.~A. Huang, H.~W. Weber, and A.~S. Bugaev.
\newblock {Nuclear magnetic resonance study of a Bi$_2$Te$_3$ topological
  insulator}.
\newblock \emph{Physics of the Solid State}, 57\penalty0 (9):\penalty0
  1741--1745, 2015.

\bibitem[Georgieva et~al.(2016)Georgieva, Rybicki, Guehne, Williams, Chong,
  Kadowaki, Garate, and Haase]{Georgieva2016}
Nataliya~M. Georgieva, Damian Rybicki, Robin Guehne, Grant V.~M. Williams,
  Shen~V. Chong, Kazuo Kadowaki, Ion Garate, and J{\"u}rgen Haase.
\newblock {$^{77}$Se nuclear magnetic resonance of topological insulator
  Bi$_2$Se$_3$}.
\newblock \emph{Physical Review B}, 93\penalty0 (19):\penalty0 195120, 2016.

\bibitem[Matano et~al.(2016)Matano, Kriener, Segawa, Ando, and
  Zheng]{Matano2016}
K.~Matano, M.~Kriener, K.~Segawa, Y.~Ando, and Guo-qing Zheng.
\newblock {Spin-rotation symmetry breaking in the superconducting state of
  Cu$_x$Bi$_2$Se$_3$}.
\newblock \emph{Nature Physics}, 12\penalty0 (9):\penalty0 852, 2016.

\bibitem[Antonenko et~al.(2017{\natexlab{a}})Antonenko, Charnaya, Nefedov,
  Podorozhkin, Uskov, Bugaev, Lee, Chang, Naumov, Perevozchikova,
  et~al.]{Antonenko2017a}
A.~O. Antonenko, E.~V. Charnaya, D.~Yu. Nefedov, D.~Yu. Podorozhkin, A.~V.
  Uskov, A.~S. Bugaev, M.~K. Lee, L.~J. Chang, S.~V. Naumov, Yu.~A.
  Perevozchikova, et~al.
\newblock {NMR studies of single crystals of the topological insulator
  Bi$_2$Te$_3$ at low temperatures}.
\newblock \emph{Physics of the Solid State}, 59\penalty0 (5):\penalty0
  855--859, 2017{\natexlab{a}}.

\bibitem[Antonenko et~al.(2017{\natexlab{b}})Antonenko, Charnaya, Nefedov,
  Podorozhkin, Uskov, Bugaev, Lee, Chang, Naumov, Perevozchikova,
  et~al.]{Antonenko2017b}
A.~O. Antonenko, E.~V. Charnaya, D.~Yu. Nefedov, D.~Yu. Podorozhkin, A.~V.
  Uskov, A.~S. Bugaev, M.~K. Lee, L.~J. Chang, S.~V. Naumov, Yu.~A.
  Perevozchikova, et~al.
\newblock {NMR study of topological insulator Bi$_2$Te$_3$ in a wide
  temperature range}.
\newblock \emph{Physics of the Solid State}, 59\penalty0 (12):\penalty0
  2331--2339, 2017{\natexlab{b}}.

\bibitem[Taylor et~al.(2012)Taylor, Leung, Lake, and Bouchard]{Taylor2012}
Robert~E. Taylor, Belinda Leung, Michael~P. Lake, and Louis-S Bouchard.
\newblock {Spin--lattice relaxation in bismuth chalcogenides}.
\newblock \emph{The Journal of Physical Chemistry C}, 116\penalty0
  (32):\penalty0 17300--17305, 2012.

\bibitem[Koumoulis et~al.(2014)Koumoulis, Leung, Chasapis, Taylor, King~Jr.,
  Kanatzidis, and Bouchard]{Koumoulis2014}
Dimitrios Koumoulis, Belinda Leung, Thomas~C. Chasapis, Robert Taylor, Daniel
  King~Jr., Mercouri~G. Kanatzidis, and Louis-S. Bouchard.
\newblock {Understanding Bulk Defects in Topological Insulators from
  Nuclear-Spin Interactions}.
\newblock \emph{Advanced Functional Materials}, 24\penalty0 (11):\penalty0
  1519--1528, 2014.

\bibitem[Levin et~al.(2016)Levin, Riedemann, Howard, Jo, Bud'ko, Canfield, and
  Lograsso]{Levin2016}
E.~M. Levin, Trevor~M. Riedemann, A.~Howard, Na~H. Jo, Sergey~L Bud'ko, Paul~C.
  Canfield, and Thomas~A. Lograsso.
\newblock {$^{125}$Te NMR and Seebeck effect in Bi$_2$Te$_3$ synthesized from
  stoichiometric and Te-rich melts}.
\newblock \emph{The Journal of Physical Chemistry C}, 120\penalty0
  (44):\penalty0 25196--25202, 2016.

\bibitem[Koumoulis et~al.(2013)Koumoulis, Chasapis, Taylor, Lake, King,
  Jarenwattananon, Fiete, Kanatzidis, and Bouchard]{Koumoulis2013}
Dimitrios Koumoulis, Thomas~C. Chasapis, Robert~E Taylor, Michael~P. Lake,
  Danny King, Nanette~N. Jarenwattananon, Gregory~A. Fiete, Mercouri~G
  Kanatzidis, and Louis-S Bouchard.
\newblock {NMR probe of metallic states in nanoscale topological insulators}.
\newblock \emph{Physical Review Letters}, 110\penalty0 (2):\penalty0 026602,
  2013.

\bibitem[Choi and Lee(2018)]{Choi2018}
Dong~Min Choi and Cheol~Eui Lee.
\newblock {$^{77}$Se Nuclear Magnetic Resonance Study of the Surface Effect in
  Topological Insulator Bi$_2$Se$_3$ Nanoparticles}.
\newblock \emph{J. Korean Phys. Soc.}, 72\penalty0 (7):\penalty0 835--837,
  April 2018.

\bibitem[Choi et~al.(2018)Choi, Lee, and Lee]{Choi2018b}
Dong~Min Choi, Kyu~Won Lee, and Cheol~Eui Lee.
\newblock {$^{125}$Te nuclear magnetic resonance and impedance spectroscopy
  study of topological insulator Bi$_2$Te$_3$ nanoparticles mixed with
  insulating Al$_2$O$_3$ nanoparticles}.
\newblock \emph{Materials Research Express}, 2018.
\newblock URL \url{http://iopscience.iop.org/10.1088/2053-1591/aaf524}.

\bibitem[Boutin et~al.(2016)Boutin, Ram{\'\i}rez-Ruiz, and Garate]{Boutin2016}
S.~Boutin, J.~Ram{\'\i}rez-Ruiz, and I.~Garate.
\newblock {Tight-binding theory of NMR shifts in topological insulators and}.
\newblock \emph{Phys. Rev. B}, 94:\penalty0 115204, September 2016.

\bibitem[Young et~al.(2012)Young, Lai, Xu, Yang, Gu, Pan, Valla, Shu, Sankar,
  and Chou]{Young2012}
Ben-Li Young, Zong-Yo Lai, Zhijun Xu, Alina Yang, G.~D. Gu, Z.-H. Pan,
  T.~Valla, G.~J. Shu, R.~Sankar, and F.~C. Chou.
\newblock {Probing the bulk electronic states of Bi$_{2}$Se$_{3}$ using nuclear
  magnetic resonance}.
\newblock \emph{Physical Review B}, 86\penalty0 (7):\penalty0 075137, 2012.

\bibitem[Nisson et~al.(2013)Nisson, Dioguardi, Klavins, Lin, Shirer, Shockley,
  Crocker, and Curro]{Nisson2013}
D.~M. Nisson, A.~P. Dioguardi, P.~Klavins, C.~H. Lin, K.~Shirer, A.~C.
  Shockley, J.~Crocker, and N.~J. Curro.
\newblock {Nuclear magnetic resonance as a probe of electronic states of
  Bi$_{2}$Se$_{3}$}.
\newblock \emph{Physical Review B}, 87\penalty0 (19):\penalty0 195202, 2013.

\bibitem[Nisson et~al.(2014)Nisson, Dioguardi, Peng, Yu, and Curro]{Nisson2014}
D.~M. Nisson, A.~P. Dioguardi, X.~Peng, D.~Yu, and N.~J. Curro.
\newblock {Anomalous nuclear magnetic resonance spectra in Bi$_{2}$Se$_{3}$
  nanowires}.
\newblock \emph{Phys. Rev. B}, 90\penalty0 (12):\penalty0 125121, September
  2014.

\bibitem[Mukhopadhyay et~al.(2015)Mukhopadhyay, Kr{\"a}mer, Mayaffre, Legg,
  Orlita, Berthier, Horvati{\'c}, Martinez, Potemski, Piot,
  et~al.]{Mukhopadhyay2015}
S.~Mukhopadhyay, S.~Kr{\"a}mer, H.~Mayaffre, H.~F. Legg, M.~Orlita,
  C.~Berthier, M.~Horvati{\'c}, G.~Martinez, M.~Potemski, B.~A. Piot, et~al.
\newblock {Hyperfine coupling and spin polarization in the bulk of the
  topological insulator Bi$_2$Se$_3$}.
\newblock \emph{Physical Review B}, 91\penalty0 (8):\penalty0 081105, 2015.

\bibitem[Blaha et~al.(2018)Blaha, Schwarz, Madsen, Kvasnicka, Luitz, Laskowski,
  Tran, and Marks]{wien2k}
P.~Blaha, K.~Schwarz, G.~K.~H. Madsen, D.~Kvasnicka, J.~Luitz, R.~Laskowski,
  F.~Tran, and L.~D. Marks.
\newblock \emph{{WIEN2k, An Augmented Plane Wave + Local Orbitals Program for
  Calculating Crystal Properties (Karlheinz Schwarz, Techn. Universit\"{a}t
  Wien, Austria)}}.
\newblock 2018.

\bibitem[MacDonald et~al.(1980)MacDonald, Picket, and Koelling]{MacDonald1980}
A.~H. MacDonald, W.~E. Picket, and D.~D. Koelling.
\newblock A linearised relativistic augmented-plane-wave method utilising
  approximate pure spin basis functions.
\newblock \emph{Journal of Physics C: Solid State Physics}, 13\penalty0
  (14):\penalty0 2675, 1980.
\newblock URL \url{http://stacks.iop.org/0022-3719/13/i=14/a=009}.

\bibitem[Perdew et~al.(1996)Perdew, Burke, and Ernzerhof]{Blaha1996}
John~P. Perdew, Kieron Burke, and Matthias Ernzerhof.
\newblock Generalized gradient approximation made simple.
\newblock \emph{Phys. Rev. Lett.}, 77:\penalty0 3865--3868, Oct 1996.
\newblock \doi{10.1103/PhysRevLett.77.3865}.
\newblock URL \url{https://link.aps.org/doi/10.1103/PhysRevLett.77.3865}.

\bibitem[Huang et~al.(2012)Huang, Chu, Kung, Lee, Sankar, Liou, Wu, Kuo, and
  Chou]{Huang2012}
F.-T. Huang, M.-W. Chu, H.~H. Kung, W.~L. Lee, R.~Sankar, S.-C. Liou, K.~K. Wu,
  Y.~K. Kuo, and F.~C. Chou.
\newblock {Nonstoichiometric doping and Bi antisite defect in single crystal
  Bi${}_{2}$Se${}_{3}$}.
\newblock \emph{Phys. Rev. B}, 86:\penalty0 081104, Aug 2012.
\newblock \doi{10.1103/PhysRevB.86.081104}.
\newblock URL \url{https://link.aps.org/doi/10.1103/PhysRevB.86.081104}.

\bibitem[Haase et~al.(1994)Haase, Conradi, Grey, and Vega]{Haase1994}
J.~Haase, M.~S. Conradi, C.~Grey, and A.~Vega.
\newblock {Population Transfers for NMR of Quadrupolar Spins in Solids}.
\newblock \emph{J. Magn. Reson. Ser. A}, 109:\penalty0 90--97, 1994.

\bibitem[Biero{\'{n}} and Pyykk{\"o}(2001)]{Bieron2001}
Jacek Biero{\'{n}} and Pekka Pyykk{\"o}.
\newblock {Nuclear Quadrupole Moments of Bismuth}.
\newblock \emph{Phys. Rev. Lett.}, 87\penalty0 (13):\penalty0 189--4, September
  2001.

\bibitem[Haase and Oldfield(1993)]{Haase1993}
J.~Haase and E.~Oldfield.
\newblock {Spin-Echo Behavior of Nonintegral-Spin Quadrupolar Nuclei in
  Inorganic Solids}.
\newblock \emph{J. Magn. Reson. Ser. A}, 101\penalty0 (1):\penalty0 30--40,
  January 1993.

\bibitem[Sobelman(2012)]{Sobelman2012}
Igor~I. Sobelman.
\newblock \emph{{Atomic Spectra and Radiative Transitions}}.
\newblock Springer Science {\&} Business Media, Berlin, Heidelberg, December
  2012.

\bibitem[K{\"o}hler and W{\"o}chner(1975)]{Kohler1975}
H.~K{\"o}hler and E.~W{\"o}chner.
\newblock {The g-factor of the conduction electrons in Bi$_2$Se$_3$}.
\newblock \emph{physica status solidi (b)}, 67\penalty0 (2):\penalty0 665--675,
  1975.

\bibitem[Wolos et~al.(2016)Wolos, Szyszko, Drabinska, Kaminska, Strzelecka,
  Hruban, Materna, Piersa, Borysiuk, Sobczak, and Konczykowski]{Wolos2016}
A.~Wolos, S.~Szyszko, A.~Drabinska, M.~Kaminska, S.~G. Strzelecka, A.~Hruban,
  A.~Materna, M.~Piersa, J.~Borysiuk, K.~Sobczak, and M.~Konczykowski.
\newblock {g-factors of conduction electrons and holes in Bi$_2$Se$_3$
  three-dimensional topological insulator}.
\newblock \emph{Physical Review B}, 93\penalty0 (1):\penalty0 155114, April
  2016.

\end{thebibliography}

\end{document}